\documentclass[conference,prl, superscriptaddress]{IEEEtran}
\IEEEoverridecommandlockouts
\usepackage[utf8]{inputenc}
\usepackage{algorithm}
\usepackage{algpseudocode}
\usepackage{fullpage}
\usepackage{amsmath,amsthm}
\usepackage{amssymb}
\usepackage{graphicx}
\usepackage{subcaption}
\usepackage[usenames,dvipsnames,table]{xcolor}
\usepackage{braket}

\renewcommand{\thesection}{\Roman{section}} 
\renewcommand{\thesubsection}{\thesection.\Roman{subsection}}
\renewcommand{\thesubsubsection}{\thesubsection.\Roman{subsubsection}}

\begin{document}

\title{Optimal Entanglement Distribution using Satellite Based Quantum Networks}

\author{\IEEEauthorblockN{Nitish K. Panigrahy\IEEEauthorrefmark{2},
Prajit Dhara\IEEEauthorrefmark{3}, Don Towsley\IEEEauthorrefmark{4}, Saikat Guha\IEEEauthorrefmark{3} and Leandros Tassiulas\IEEEauthorrefmark{2}}

\IEEEauthorblockA{\IEEEauthorrefmark{2}Department of Electrical Engineering and Institute for Network Science, Yale University, CT, 06511, USA\\ \IEEEauthorrefmark{3}Wyant College of Optical Sciences, University of Arizona, Tucson, AZ, 85719, USA\\ \IEEEauthorrefmark{4}University of Massachusetts, Amherst, MA, 01003, USA \\
Email: \IEEEauthorrefmark{2}nitishkumar.panigrahy@yale.edu, leandros.tassiulas@yale.edu \\ \IEEEauthorrefmark{3}prajitd@email.arizona.edu, saikat@optics.arizona.edu, \IEEEauthorrefmark{4}towsley@cs.umass.edu}}
\maketitle

\begin{abstract}
Recent technological advancements in satellite based quantum communication has made it a promising technology for realizing global scale quantum networks. Due to better loss distance scaling compared to ground based fiber communication, satellite quantum communication can distribute high quality quantum entanglements among ground stations that are geographically separated at very long distances. This work focuses on optimal distribution of bipartite entanglements to a set of pair of ground stations using a constellation of orbiting satellites. In particular, we characterize  the optimal satellite-to-ground station transmission scheduling policy with respect to the aggregate entanglement distribution rate subject to various resource constraints at the satellites and ground stations. We cast the optimal transmission scheduling problem as an integer linear programming problem and solve it efficiently for some specific scenarios. Our framework can also be used as a benchmark tool to measure the performance of other potential transmission scheduling policies.

\end{abstract}

\maketitle

\section{Introduction}
Quantum entanglement distribution over long distances is necessary for numerous quantum applications such as quantum key distribution (QKD) \cite{Shor00}, teleportation \cite{Ma12}, and distributed quantum computing \cite{Broadbent09}. Two most promising solutions in which entanglements can be distributed over global distances are, ground based fiber connection with a chain of quantum repeaters, and satellite based free space connection.
In a ground-based fiber connection, entangled photonic qubits are sent through a single optical fiber link connecting two remote users. It is well known that in such a direct connection, the transmissivity and hence the entanglement distribution rate decays exponentially with the distance between the remote users \cite{Takeoka14}, \cite{Pirandola17}. Thus in practice, photonic qubits can only be distributed up to distances of a few hundred kilometers in fibers using direct connection. Further long distance entanglement can be achieved by inserting a chain of quantum repeaters between the two remote users. Quantum repeaters \cite{Dur99} equipped with quantum memories \cite{Lvovsky09} perform entanglement swapping and entanglement purification \cite{Dur99} to combine elementary repeater-to-repeater link level entanglements providing high quality end-to-end user entanglements. 

An alternative strategy to achieve long distance entanglement is to use satellite based free space connection. In the case of optical satellite links, the majority of photon's propagation path is in empty space, thus incurring much less channel loss and decoherence as compared to a ground based fiber link. Also, recent experimental successes \cite{Liao17, Ren17} in satellite based quantum communication, have made it possible to envision a global scale satellite based quantum network.

In a satellite based quantum network, satellites are deployed in constellations. The goal is to generate entanglements between a set of pairs of ground stations. While both down-link and up-link entanglement distribution architectures have been proposed, the latter has an additional loss component due to early atmospheric diffraction also known as the shower-curtain effect \cite{Sidhu20}. In the double down-link distribution architecture, satellites and ground stations act as transmitters and receivers respectively. Each satellite holds entanglement generating photon sources that create pairs of entangled photons and distribute the pairs to ground stations. 

Recently, Khatri et al. \cite{Khatri21}, provided a detailed analysis of the double down-link architecture for a satellite constellation based on polar orbits. The authors studied multiple satellite configurations with a goal to minimize the total number of deployed satellites and maximize overall entanglement distribution rates across a set of pair of ground stations. Assuming a time discretized system, at the beginning of each time epoch, the authors propose to schedule the satellite-to-ground station transmissions in a greedy non-optimal manner. Also, the authors ignore resource constraints in their formulation. To the best of our knowledge, the problem of optimal scheduling of satellite-to-ground station transmissions in a satellite based quantum network has not been looked at in the literature.

Resources in satellite based quantum networks, such as number of transmitters per satellite, photon sources and number of receivers, are limited. Therefore, effective allocation of network resources and scheduling transmissions are important to maximize the overall performance of the network for a given constellation of satellites and set of  ground stations. This work serves for this purpose. To the best of our knowledge, this is the first work that aims at characterizing the optimal transmission scheduling policy for a satellite based quantum network. In this work, we aim to assign satellites to ground station pairs dynamically based on the current state information of the satellites and ground stations along with their associated entanglement generation rates and fidelities of entanglements. Consequently, our framework can be used as a benchmark tool for other potential transmission scheduling policies.

\section{Technical Preliminaries}
In this section, we introduce the system model used in the rest of the paper.
\begin{figure*}[!htbp]
\centering
\begin{minipage}{0.5\textwidth}
\includegraphics[width=1\textwidth]{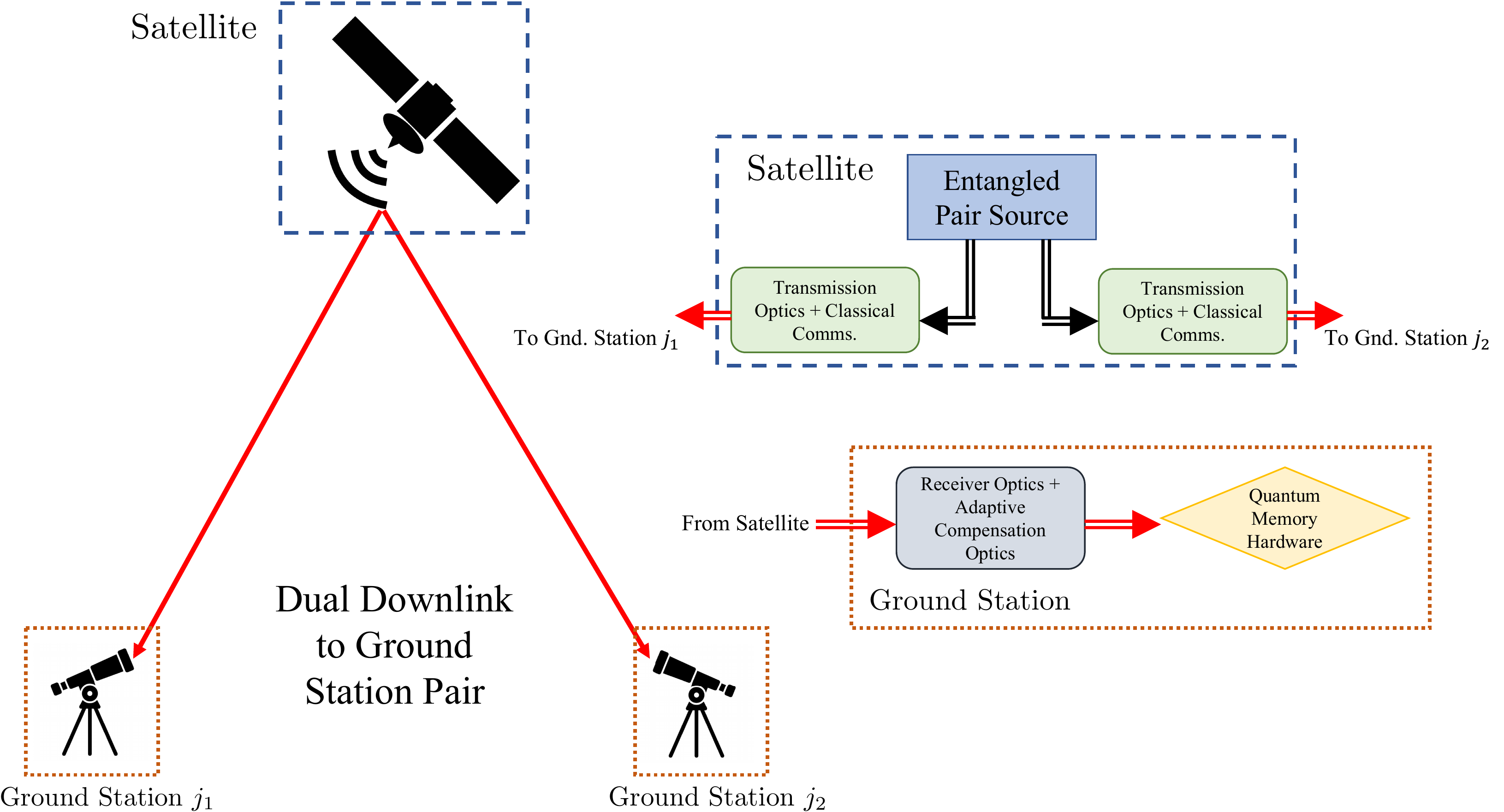}
\subcaption{}
\end{minipage}
\begin{minipage}{0.45\textwidth}
\includegraphics[width=1\textwidth]{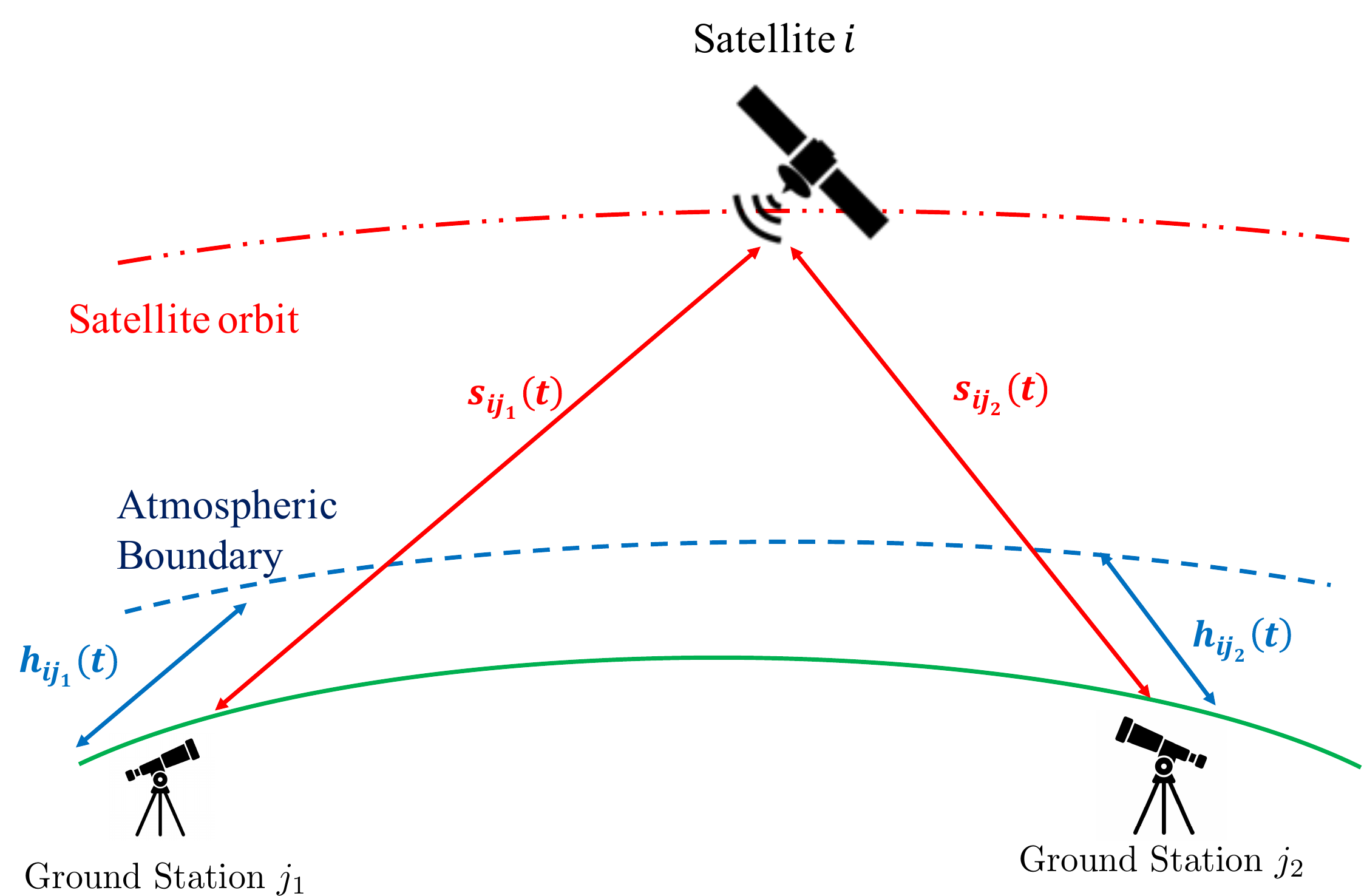}
\subcaption{}
\end{minipage}\hfill
\caption{Dual downlink architecture for photonic entanglement distribution. (a) The satellite platform consists of spontaneous parametric down conversion (SPDC) based entangled pair sources that produce the state in Eq.~\eqref{eqn:srcnative} and distribute each qubit to the ground stations, with the necessary transmission optics. The ground stations consists of receiver optics and adaptive optics (to minimize atmospheric distortion) that couple into the quantum memory hardware. We assume that the system has perfect timing synchronization, accurate pointing and tracking, and inter-station terrestrial classical communication links that are required for the protocol to succeed. (b) Total (Free space + atmospheric) and atmospheric distances of Satellite-to-ground station channels at time $t$.}
\label{shot-noise-tracestat-sat}
\end{figure*}
\subsection{Satellite constellations and ground station pairs}\label{subsec:constellation}
We consider a satellite based quantum network that distributes shared entangled photonic quantum states between a set of ground station pairs. Denote $S$, $G$, and $F$ as the set of satellites, ground stations, and pairs of ground stations that request entanglement respectively. The ground stations are located on earth, which rotates around its axis with a rotation period of $24$ hours. The satellites are deployed in satellite constellations and orbit around the earth at certain altitudes. Several examples of constellations that has been considered for classical and quantum communications include Polar, Walker, Iridium, Starlink, and Kuiper \cite{Khatri21}. While our results are applicable to any constellation, for simulation purpose, we will use the polar constellation. A complete simulation study and comprehensive comparison of other constellations remains part of our future work.

In a satellite based quantum network, each satellite has photon sources that produce entangled pair of photons. Satellites send each of the photons to a pair of ground stations using a down-link channel. After an entangled pair is successfully received by each of the two ground stations, it can be used by any quantum application such as QKD or quantum teleportation.

We denote the elevation angle limit to be $\theta_e$. Thus, the entangled photon can successfully be received at the ground station as long as the elevation angle (angle between the satellite and horizon at the ground station) of a satellite from the horizon exceeds $\theta_e$. The elevation angle is used to account for terrain obstructions, such as buildings and mountains. For successful entanglement distribution, the elevation angles for a satellite from both ground stations must exceed $\theta_e$.

We also assume time is divided into fixed length time slots. Both the satellites and ground stations have limited number of resources, such as number of photon sources and receivers. Thus at the beginning of each time slot, based on current positions of the satellites and the ground stations, the satellites  are allocated to ground station pairs to maximize the overall system-wide entanglement distribution rate subject to resource constraints. We discuss this in detail in Section \ref{sec:probform}.

\subsection{Loss and Noise model}
Entanglement distribution over satellite links rely heavily on utilizing photonic entangled pair generation source and free-space optical communication strategies. The location and configuration these sources, links and any associated hardware depends on the quantum link configuration chosen by the user(s). 

For the dual downlink configuration highlighted in Section II.A, we utilize a spontaneous parametric down-conversion based dual-rail polarization entangled pair source. This source is well-studied and widely utilized for other tasks that involve quantum entanglement generation~\cite{Krovi2016-my,Kok2000-ml}. The output quantum state from such a source is described in the photon number Fock basis (upto the support of $2$ photon pairs)~\cite{Kok2000-ml}:

\begin{align}
		\ket{\psi^\pm}&\!=   N_0 \left[\!\sqrt{p(0)} \ket{0,0;0,0}\!  +\! \sqrt{\frac{p(1)}{2}} \left(\ket{1,0;0,1}\pm\ket{0,1;1,0} \right) \right.\nonumber\\
		&\left. +  \sqrt{\frac{p(2)}{3}}\left(\ket{2,0;0,2}\pm \ket{1,1;1,1}+\ket{0,2;2,0}\right)\right],
		\label{eqn:srcnative}
\end{align}
where $N_0$ is a normalization constant and the term coefficients $ p(n) $ are given by,
\begin{equation}
p(n) = (n+1)\frac{{N_s}^n}{(N_s + 1)^{n+2}}. 
\label{eqn:geometric_dist}
\end{equation}
with $N_s$ denoting the mean photon number per mode of the state. The entangled pair (in the dual rail basis) component of the above state is $\left(\ket{1,0;0,1}\pm\ket{0,1;1,0} \right) \sqrt{2}$; the vacuum portion is $\ket{0,0;0,0}$, and all the other terms are spurious two-pair emission terms. By controlling the $N_s$ parameter one can limit the emission of these two-pair terms whilst making sure the output state has a considerable amount of the utilizable entanglement~\cite{Krovi2016-my}. 

Since the link involves free-space optical transmission, any analysis of such links must take into account the characteristics of the optical channel. For the present study, we account for transmission loss by studying the property of the state after it passes through a bosonic pure loss channel. Each qubit (i.e. pair of modes) undergoes loss that is parametrized by a channel transmissivity $ \eta_{ig}(t) $ for a channel between satellite $ i $ and ground station $g$ at time $t$. Details of the $\eta_{ig}(t) $ calculation  is given in Section II.C. 

In general, transmission through a pure loss channel reduces the mean photon number of the transmitted state, and result in a pure state becoming mixed. Hence, this affects the probability that both qubits of the entangled pairs reach both parties successfully. Additionally, since the output state has two-pair terms, loss may cause these terms to lose a single photon and resemble the entangled pair for each party. Note, that such terms are not necessarily entangled, and hence result in a degradation of the fidelity (to the ideal Bell pair) of a distributed pair. Details of the model and the mathematical analysis of the state is given in Ref.~\cite{Dhara2021-ch}.

Additionally, our analysis takes into account excess noise in the optical links (in the form of unfiltered background photons) and dark clicks  in the detectors of the associated hardware. The inclusion of background photons occurs at random and can only be detected when the users detect their individual states. However since the output quantum state is a superposition of multiple terms, the user will be unable to distinguish a background photon from photons generated by the source. As a simple example, in the absence of loss, the detection of the state $\ket{0,1}$ by a user could either mean they received a qubit (of the generated ebit), or they detected the vacuum state with one excess photon (which is not entangled). Hence given a detection pattern, the output state description is best expressed as a statistical mixture of the possible states that could have given rise to the pattern. This limits the fidelity of the output state. Paired with a lossy optical transmission, such a noise process can be highly detrimental to the quality of the shared entanglement. This is simply because, with higher loss there are fewer qubits that are successfully transmitted. Since noise is independent of loss, there would be a threshold after which background photons significantly eclipse the entangled photons and drive the fidelity of the distributed state below a manageable/suitable threshold. In the present work, we take into account this effect implicitly, by working in loss and noise regimes that remain above the desired fidelity threshold. A detailed analysis is given in Appendix D of  Ref.~\cite{Dhara2021-ch}.


\subsection{Channel Transmissivities}\label{sub-trans}
The transmissivity $ \eta_{ig}(t) $ for a channel between satellite $i$ and ground station $g$ at time $t$ depends on the free space and atmospheric distances as shown in Figure \ref{shot-noise-tracestat-sat} (b). Let $s_{ig}(t)$ be the distance between satellite $i$ and ground station $g$ at time $t$. Denote $h_{ig}(t)$ be the distance between ground station $g$ and atmospheric boundary when connected to satellite $i$. We consider optical links with circular apertures of diameter $d_{iT}$ and $d_{gR}$ for the  transmitter and receiver telescopes at satellite $i$ and ground station $g$ respectively, operating at some wavelength $\lambda$. The free space, atmospheric and the overall channel transmissivity for such a link is well approximated by, 
\begin{align}
    &\eta_{ig}^{(f)}(t) =\frac{(\pi d_{iT}^2/4)(\pi d_{gR}^2 /4)}{(\lambda s_{ig}(t))^2},\quad \eta_{ig}^{(a)}(t) = e^{-\alpha h_{ig}(t)}\nonumber\\
    &\eta_{ig}(t) =\eta_{ig}^{(f)}(t)\eta_{ig}^{(a)}(t),
\end{align}
\noindent where $\alpha$ is the atmospheric extinction coefficient.
\begin{table}[!htbp]
\begin{center}
		\begin{tabular}{ l|l} 
			\hline
			$S$&Set of satellites\\
			$G$&Set of ground stations\\
			$F$&Set of pair of ground stations\\
			$N$&Number of satellites\\
			$M$&Number of pairs of ground stations\\
			$R_g$&Maximum number of receivers at ground station $g$\\
			$T_i$&Maximum number of transmitters at satellite $i$\\
			$L_j$&Maximum number of simultaneous connections allowed \\
			&for ground station pair $j$\\
			$\theta_e$&Elevation angle limit (from the horizon)\\
			$\kappa$& Total number of time slots\\
			$\Delta$&Slot duration (in seconds)\\
			$t$& A time slot ($1,\cdots, \kappa$)\\
			$e_{ig}(t)$& Elevation angle of satellite $i$ from ground station\\
			&$g$ at time $t$\\
			$F^{th}$&Fidelity threshold\\
			$x_{ij}(t)$&Optimization variable denoting if satellite $i$ is \\
			&assigned to ground station pair $j$ at time $t$\\
			$w_{ij}(t)$&Weight associated with the assignment of satellite $i$\\
			&to ground station pair $j$ at time $t$\\
			$\psi_{ij}(t)$ & Entanglement rate associated with satellite $i$ for \\
			&entanglement of the ground station pair $j$ at time $t$\\
			$\chi_{ij}(t)$ & Fidelity associated with satellite $i$ for entanglement\\
			&of the ground station pair $j$ at time $t$\\
			$C_{ij}(t)$& Indicator variable denoting if satellite $i$ covers both \\
			&ground stations and $\chi_{ij}(t) \ge F^{th}$ in pair $j$ at time $t$\\
			\hline
		\end{tabular}
\end{center}		
		\caption{Summary of Notations.}
		\label{table:smrnd}
\end{table}
\vspace{-0.5cm}
\section{Problem Formulation}\label{sec:probform}
We now propose the optimal transmission scheduling policy associated with satellite based quantum network. We summarize the notations used in Table \ref{table:smrnd}. 

We assume time is slotted with slot duration $\Delta$ seconds. Let $\kappa$ denote the total number of time slots. Let $t$ be any arbitrary time slot with $t = 1, \cdots, \kappa$. Let $x_{ij}(t)$ be the binary decision variable denoting if satellite $i$ is assigned to create entanglement between ground station pair $j$ at time $t.$ In any transmission scheduling policy, the overall goal is to assign $S$ to $F$ so as to maximize the objective $O(t) = \sum_{i\in S}\sum_{j\in F}\omega_{ij}(t)C_{ij}(t)x_{ij}(t).$ Here, $\omega_{ij}(t)$ denotes the weight/utility associated with the assignment of satellite $i$ to ground station pair $j$. Possible choices of $\omega_{ij}(t)$ include the entanglement generation rate $\psi_{ij}(t)$ or the average arrival rate of requests for creating entanglement between ground stations in pair $j$ for a particular quantum application such as: QKD or quantum teleportation. Let $j = \{j_1, j_2\}$, i.e. pair $j$ consists of ground stations $j_1$ and $j_2.$ Let $C_{ij}(t)$ be the indicator variable defined as follows.
\begin{equation}\label{eq:connection-var}
    C_{ij}(t)=
\begin{cases}
     1,& \chi_{ij}(t) \ge F^{th} \text{ and } e_{ij_1}(t), e_{ij_2}(t) \ge \theta_e ;\\
     0,& \text{otherwise},
\end{cases}
\end{equation}
Here, $\chi_{ij}(t)$ denotes the fidelity of the entangled state between satellite $i$ and pair $j$ with $F^{th}$ being the minimum fidelity threshold. $e_{ij_1}(t)$ and $e_{ij_2}(t)$ are the elevation angles of satellite $i$ from ground stations $j_1$ and  $j_2$ respectively. Note that, $\psi_{ij}(t)$ and  $\chi_{ij}(t)$ are determined at the beginning of time slot $t$ according to the noise and loss models discussed in Section II.B.

Thus the optimal transmission scheduling problem (OPT-SAT) at time slot $t$ can be cast as the following integer programming problem.
\begin{subequations}\label{eq:sat-opt}
\begin{align}
\text{\bf{OPT-SAT:}} 
\max \quad&\sum_{i\in S}\sum_{j\in F} \omega_{ij}(t)C_{ij}(t) x_{ij}(t)\label{eq:stat-sat1}\displaybreak[0]\\
\quad&\sum_{i\in S}\sum_{j\in F, g\in j} x_{ij}(t) \le R_g, \forall g\in G,\label{eq:stat-sat2}\\
\quad&\sum_{j\in F} x_{ij}(t) \le T_i, \forall i \in S\label{eq:stat-sat3}\\
\quad&\sum_{i\in S} x_{ij}(t) \le L_j, \forall j \in F\label{eq:stat-sat4}\\
\text{s.t.} \quad &x_{ij}(t)\in \{0,1\} \quad \forall i \in S, \forall j\in F,\label{eq:stat-sat4}
\end{align}
\end{subequations}
\noindent where $T_i$ and $R_g$ are the maximum number of transmitters and receivers at  satellite $i$ and ground station $g$ respectively. $L_j$ denotes the maximum number of simultaneous connections allowed for pair $j$. 
\begin{figure}[htbp]
\begin{minipage}{0.3\textwidth}
\includegraphics[width=0.85\textwidth, height=0.45\textwidth]{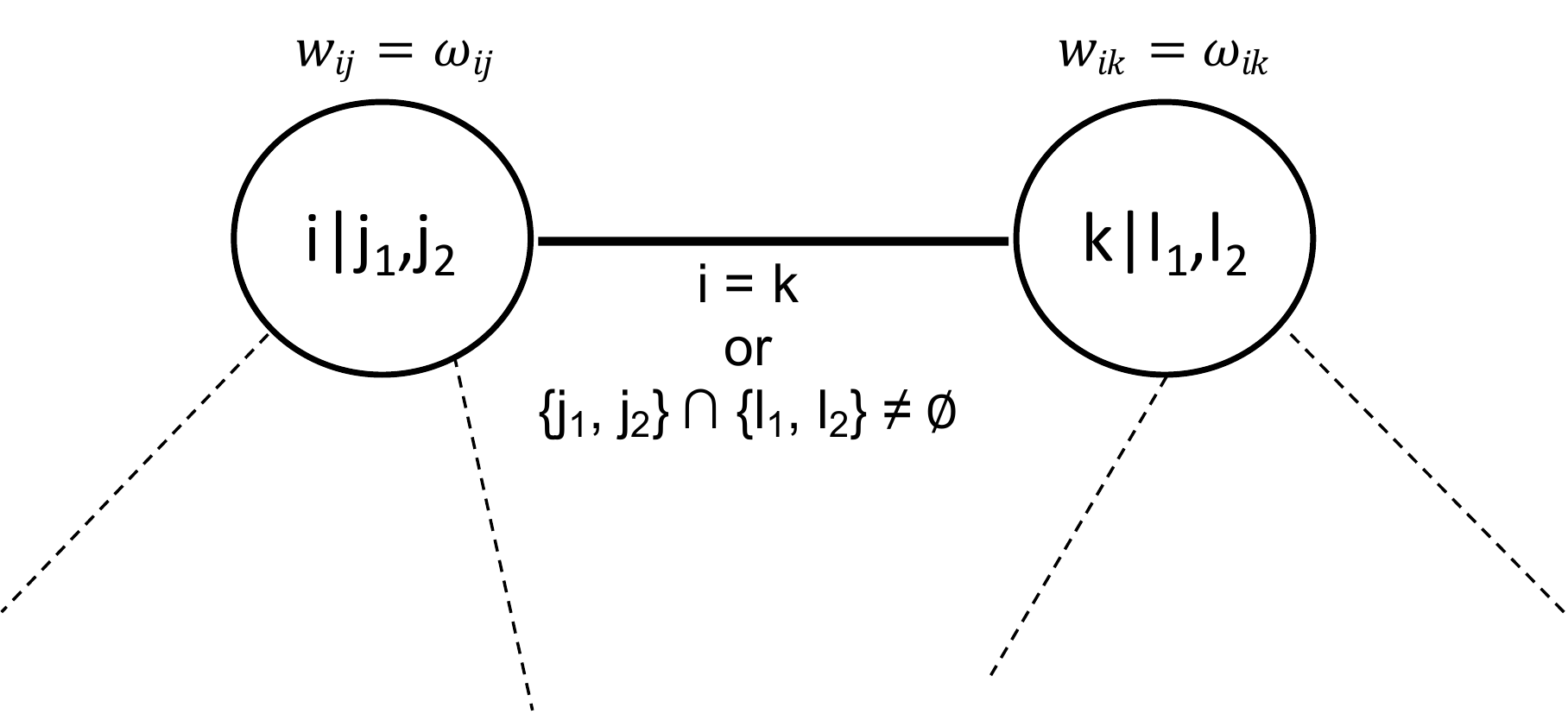}
\subcaption{}
\end{minipage}
\begin{minipage}{0.3\textwidth}
\includegraphics[width=0.5\textwidth, height=0.8\textwidth]{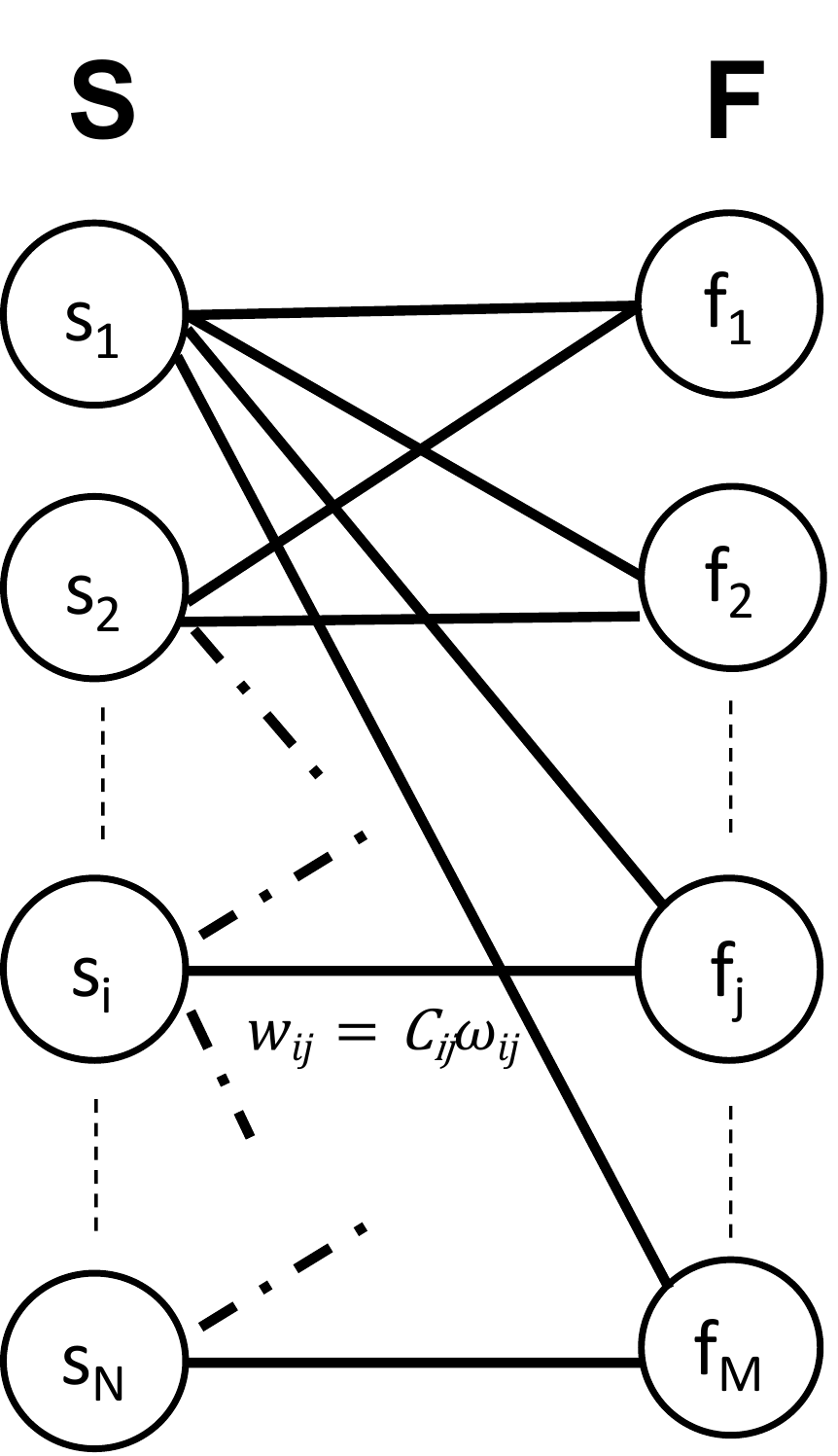}
\subcaption{}
\end{minipage}
\caption{Solving OPT-SAT can be equivalent to (a) solving a maximum weight independent set problem when $R_g = 1, T_i=1, L_j = 1$ and (b) solving a matching problem when $R_g = N, T_i=1, L_j = 1$.}
\vspace{-0.5cm}
\label{fig:graphs}
\end{figure}

Depending on request requirements, a ground station $g$ can be part of multiple ground station pairs and thus is not allowed  to be allocated to more than $R_g$ satellites. Constraint \eqref{eq:stat-sat2} corroborates the same. Constraint \eqref{eq:stat-sat3} ensures that satellite $i$ does not get allocated to more than $T_i$ ground station pairs. Similarly, Constraint \eqref{eq:stat-sat4} does not allow ground station pair $j$ to be allocated to more than $L_j$ satellites. We also assume $R_g \ge L_j, \forall j\in F, g \in j$. Constraint \eqref{eq:stat-sat4} ensures that $x_{ij}(t)$ are binary decision variables.

The optimization problem \eqref{eq:sat-opt} is an integer linear programming problem. Below we consider few specific scenarios and in some cases, design efficient exact algorithms to solve OPT-SAT. 
\vspace{-0.2cm}
\subsection{Single transmitter and single receiver scenario ($R_g = 1, T_i=1, L_j = 1$)}
We first consider the case when each satellite has only a single transmitter and each ground station has a single receiver. We can construct a graph $G_t(V, E)$ as shown in Figure  \ref{fig:graphs} (a). We set $V = \{(i|j_1,j_2), \forall i \in S, j = \{j_1, j_2\} \in F, C_{ij}(t) = 1\}.$ We assign each vertex $v \in V$ a weight $w_v = \omega_{ij}(t)$ where $v = (i|j_1,j_2)$ and $j = \{j_1, j_2\}.$ Also we define $E = \{e = <(i|j_1,j_2), (k|l_1,l_2)> \text{ s.t. } i=k \text{ or } \{j_1,j_2\} \cap \{l_1,l_2\} \ne \Phi\}.$ It can be easily seen that solving OPT-SAT in this case corresponds to obtaining the maximum weight independent set of $G_t.$

Finding the maximum weight independent set of a general graph is NP-hard. However, for many classes of graphs, a maximum weight independent set may be found in polynomial time. Some examples include claw-free graphs, P5-free graphs, perfect graphs and chordal graphs \cite{Nobili15}. Unfortunately, through careful choice of values for $C_{ij}(t)$, one can construct scenarios for which $G_t$ is not claw-free or perfect or chordal or P5-free, thus advocating for the possibility that existence of an efficient exact algorithm to solve OPT-SAT is unlikely. 

Numerous approximation algorithms have been purposed to solve the maximum weight independent set problem. Below we briefly discuss one such algorithm \cite{Kako09}. Let $\delta_v$ be the degree of vertex $v.$ We assume the initial independent set to be empty. Let $\gamma_v = \sum_{u|(u,v)\in E} w_u/w_v$ be the weighted degree of vertex $v$. Now, we add the vertex with the smallest value of  $\gamma_v$ to the independent set. We then delete this vertex and all of its neighbors from the graph. We redo this process for the remaining sub-graph until it becomes empty. Let $\rho_t$ be the approximation ratio for this algorithm. It has been shown that \cite{Kako09} $\rho_t \ge \frac{1}{\gamma+1},$ where $\gamma = \sum_{v\in V}w_v\delta_v/\sum_{v\in V}w_v$. 


\subsection{Single transmitter and multiple receiver scenario ($R_g = N, T_i=1, L_j = 1$)}
Next we consider the case when each satellite has only a single transmitter and each ground station has multiple receivers. If we are restricted to assign only one satellite to one ground station pair, then solving OPT-SAT can be transformed into finding maximum weight matching in a bipartite graph $G_t(V, E)$ with bi-partition $(S, F)$ and $V = S\cup F$ as shown in Figure \ref{fig:graphs} (b). In this case, we define a weight function on the edges $w: E \rightarrow \mathbb{R}$ as follows. We set $w_{ij} = \omega_{ij}(t)C_{ij}(t)$ for all $i \in S, j \in F.$ The maximum weight bipartite matching problem on $G_t$ can be solved efficiently by the Hungarian algorithm with a time complexity of $O(n^3)$ where $n = \max\{N, M\}$.

Note that, the case with $R_g = N, T_i\ge 1, L_j \ge 1$ can easily be efficiently solved by creating $T_i$ and $L_j$ copies of satellite $i$ and ground station pair $j$ respectively at prescribed locations and by finding the optimal bipartite matching on this extended bipartite graph.
\section{Numerical Results}

\begin{figure}[htbp]
\centering
\vspace{-0.4cm}
\includegraphics[width=0.5\textwidth, height = 0.35\textwidth]{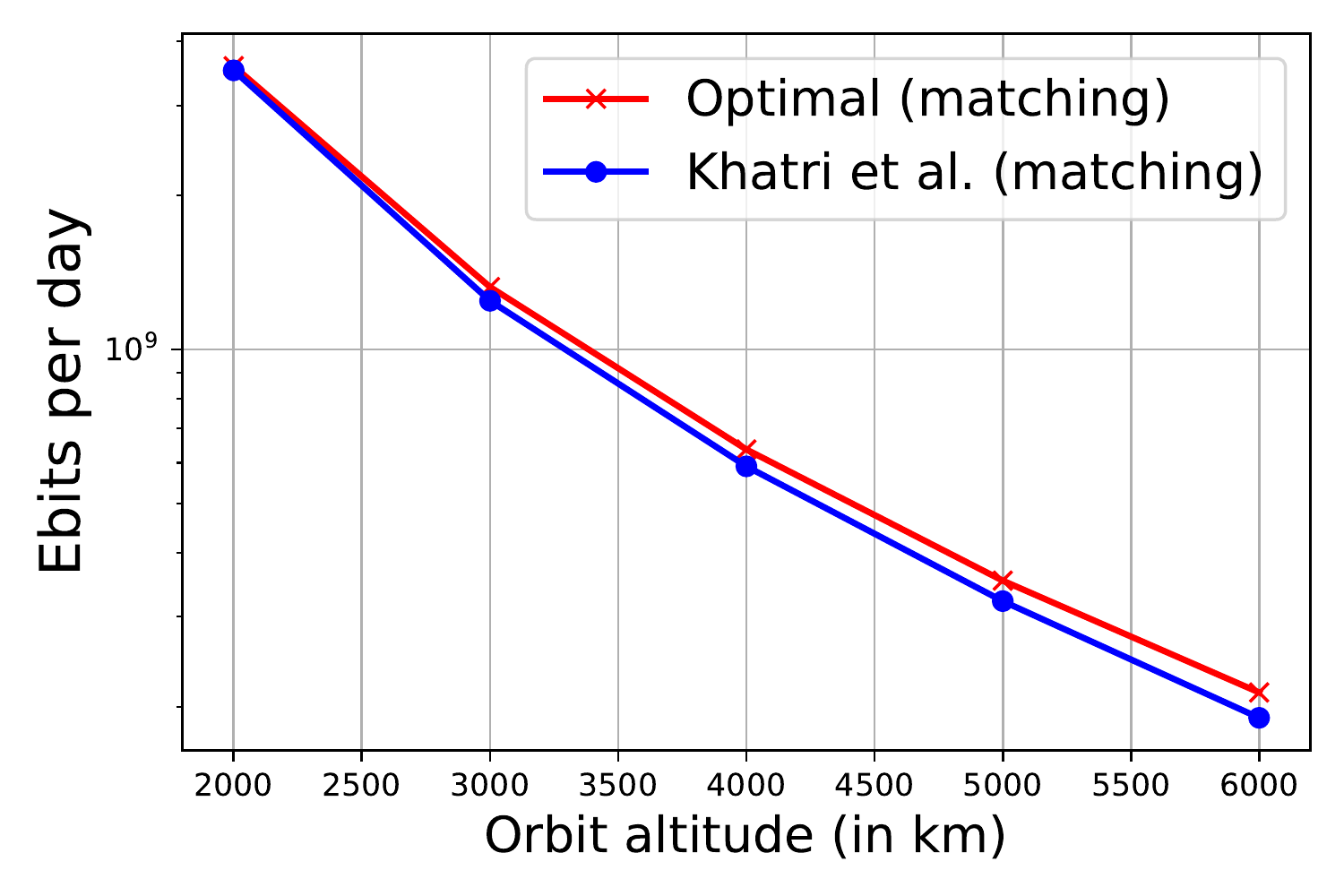}
\caption{Performance comparison of OPT-SAT with greedy scheduling policy proposed by Khatri et al. \cite{Khatri21}.}
\label{fig:sim}
\end{figure}
In this section, we use simulations to compare the performance of OPT-SAT to that of a greedy scheduling policy proposed in the literature \cite{Khatri21}. We assume polar satellite constellation where satellites are deployed in $10$ equally spaced rings with $10$ equally spaced satellites in each ring at the same orbit altitudes. More details on this constellation can be found in \cite{Khatri21}. We assume the ground station pairs to be all possible pair combinations of the cities: Toronto, New York City, London, Singapore, Sydney, Rio de Janeiro, and Mumbai. Thus we have $N = 100, M = 21.$ We set $\theta_e = 0, \Delta = 1, \kappa = 86400\; (1\text{ day})$. Again, we set $F^{th} = 0.95, d_{iT} = 0.2 m, d_{gR} = 2 m, \lambda = 737 nm, \alpha = 0.028125,$ and satellite pump power to $0.0078$. We consider a $10GHz$ repetition rate for the entanglement source, i.e. during a time slot of $1$ second, $10^{10}$ entangled photons are produced at the source. For each time slot $t = 1, \cdots, \kappa$, we consider the case when $R_g = N, T_i=1, L_j = 1$ and solve OPT-SAT by finding the maximum weight bipartite matching as described in Section III.B.

We present the results of our simulation in Figure \ref{fig:sim}. We study the effect of satellite orbit altitude on the number of successful entanglements generated across all pairs of ground stations aggregated over all time slots. Clearly, by increasing orbit altitude, the entanglement generation rate decreases. At higher altitudes, the satellite to ground station connectivity increases. I.e. the pass time or the total time for which a satellite is connected to both ground stations is more as compared to the pass time for the same satellite orbiting in lower altitudes. However, the transmission losses are huge at higher altitudes due to greater free space distances travelled by the entangled photons. Eventually, the loss component dominates and we get lower entanglement generation rate at higher altitudes. Moreover, as expected, OPT-SAT performs better compared to the state-of-the-art greedy scheduling policy. Also, notice that, the performance difference between OPT-SAT and greedy scheduling policy increases with increase in orbit altitude. At higher altitudes, a satellite covers more ground stations and a ground station is covered by more number of satellites. Thus, the feasibility region/space of the OPT-SAT problem increases with altitude and it increasingly performs better than the greedy policy. At an altitude of $6000$ kilometers, there is around $12\%$ improvement in performance of OPT-SAT compared to greedy. 
\vspace{-0.1cm}
\section{Conclusion}
In this work, we optimally scheduled the satellite-to-ground station transmissions of bipartite entangled photons in satellite based quantum network. We cast the optimal scheduling problem as an integer programming problem and discussed two specific scenarios representing different resource constraints. Using numerical simulations, we compared the performance of our proposed optimal scheduling policy to a state-of-the-art greedy scheduling policy for polar satellite constellation. Going further, we aim to extend our analysis to characterize optimal scheduling policy for multi-partite entanglement distribution. A simulation study to compare and contrast the performance of different satellite constellations remain one of our future works.
\vspace{-0.1cm}
\section{Acknowledgments}
This research was supported by the NSF Engineering Research Center for Quantum Networks (CQN), awarded under cooperative agreement number 1941583 and by NSF grant CNS-1955834.


\vspace{-0.1cm}
\begin{thebibliography}{00}

\bibitem{Shor00} P. W. Shor and J. Preskill, ``Simple proof of security of the bb84 quantum key distribution protocol," Phys. Rev. Lett., vol. 85, no. 2, p. 4-41, 2000.




\bibitem{Ma12} X. S. Ma, T. Herbst, T. Scheidl, D. Wang, S. Kropatschek, W. Naylor, B. Wittmann, A. Mech, J. Kofler, E. Anisimova et al., ``Quantum teleportation over 143 kilometres using active feed-forward," Nature, vol. 489, no. 7415, p. 269, 2012.

\bibitem{Broadbent09} A. Broadbent, J. Fitzsimons, and E. Kashefi, ``Universal Blind Quantum Computation," IEEE Symposium on Foundations of Computer Science, pp. 517-526, 2009.

\bibitem{Takeoka14} M. Takeoka, S. Guha, and M. M. Wilde, ``Fundamental rate-loss tradeoff for optical quantum key distribution," Nat. Commun.,  vol. 5, no. 5235, 2014.

\bibitem{Pirandola17} S. Pirandola, R. Laurenza, C. Ottaviani, and L. Banchi, ``Fundamental limits of repeaterless quantum communications,"  Nat. Commun., vol. 8, no. 15043, 2017.

\bibitem{Dur99} W. Dur, H.-J. Briegel, J. I. Cirac, and P. Zoller, ``Quantum repeaters based on entanglement purification," Phys. Rev. A., vol. 59, pp. 169-181, 1999.

\bibitem{Lvovsky09} A. Lvovsky, B. Sanders, and W. Tittel, ``Optical quantum memory," Nature Photonics, vol. 3, pp. 706-714, 2009.


\bibitem{Khatri21} S. Khatri, A. J. Brady, R. A. Desporte, M. P. Bart, and  J. P. Dowling, ``Spooky action at a global distance: analysis of space-based entanglement distribution for the quantum internet," npj Quantum Inf, vol. 7, no. 4, 2021.

\bibitem{Dhara2021-ch} P. Dhara, S. Johnson, C. Gagatsos, P. Kwiat, \& S. Guha,  ``Heralded-Multiplexed High-Efficiency Cascaded Source of Dual-Rail Polarization-Entangled Photon Pairs using SPDC,'' arXiv preprint, arXiv:2107.14360 [quant-ph], 2021.
\bibitem{Krovi2016-my}H. Krovi, S. Guha,  Z. Dutton,  J. Slater, C. Simon, \& W. Tittel `` Practical quantum repeaters with parametric down-conversion sources,'' {\em Appl. Phys. B}. \textbf{122}, 2016.

\bibitem{Kok2000-ml}P. Kok, \& S. Braunstein, ``Postselected versus nonpostselected quantum teleportation using parametric down-conversion,''{\em Phys. Rev. A}. \textbf{61}, 042304, 2000.



\bibitem{Liao17} S. K. Liao, W. Q. Cai, W. Y. Liu et al., ``Satellite-to-ground quantum key distribution," Nature, vol. 549, pp. 43-47, 2017.

\bibitem{Ren17}  J. G. Ren, P. Xu, H. L. Yong et al., ``Ground-to-satellite quantum teleportation," Nature, vol. 549, pp. 70-73, 2017.


\bibitem{Sidhu20} J. S. Sidhu, T. Brougham, D. McArthur, R. G. Pousa  and D. K. L. Oi, ``Finite key effects in satellite quantum key distribution," arXiv 2012.07829, 2020.

\bibitem{Nobili15} P. Nobili, and A. Sassano, ``An $O(n^2 \log n)$ algorithm for the weighted stable set problem in claw-free graphs," arXiv:1501.05775, 2015.

\bibitem{Kako09} A. Kako, T. Ono, T. Hirata, and M. M. Halldorsson, ``Approximation algorithms for the weighted independent set problem in sparse graphs," Discrete Applied Mathematics, vol. 157, pp. 617-626, 2009.

\end{thebibliography}
\end{document}